\documentclass[%
 reprint,
 superscriptaddress,
 amsmath,amssymb,
 aps,https://www.overleaf.com/project/5d5ae32d4618dd37df01b4f9
 pra,
floatfix,
]{revtex4-1}
\usepackage{amsmath}
\usepackage{lipsum}
\DeclareMathOperator{\sinc}{sinc}
\usepackage{multirow}
\usepackage{graphicx}
\usepackage{dcolumn}
\usepackage{bm}
\usepackage[colorlinks,linkcolor=blue,citecolor = blue,urlcolor=blue]{hyperref}
\usepackage{booktabs}
\usepackage{float}
\usepackage{csquotes}
\usepackage{abraces}

\usepackage{commath}
\usepackage{ulem}
\usepackage{braket}
\usepackage{color}
\usepackage{cleveref}
\usepackage{array}
\usepackage{caption}
\begin{document}

\title{Enhancing the purity of single photons in parametric down-conversion through simultaneous pump-beam and crystal-domain engineering}

\author{Baghdasar Baghdasaryan}
\email{baghdasar.baghdasaryan@uni-jena.de}
\affiliation{Theoretisch-Physikalisches Institut, Friedrich-Schiller-Universit\"at Jena, 07743 Jena, Germany}
\affiliation{Helmholtz-Institut Jena, 07743 Jena, Germany}


\author{Fabian Steinlechner}
\email{fabian.steinlechner@uni-jena.de}
 \affiliation{Fraunhofer Institute for Applied Optics and Precision Engineering, 07745 Jena, Germany}
 \affiliation{Abbe Center of Photonics, Friedrich-Schiller-University Jena, 07745 Jena, Germany}

\author{Stephan Fritzsche}
\affiliation{Theoretisch-Physikalisches Institut, Friedrich-Schiller-Universit\"at Jena, 07743 Jena, Germany}
\affiliation{Helmholtz-Institut Jena, 07743 Jena, Germany}
 \affiliation{Abbe Center of Photonics, Friedrich-Schiller-University Jena, 07745 Jena, Germany}
\date{\today}

\begin{abstract}
Spontaneous parametric down-conversion (SPDC) has shown great promise in the generation of pure and indistinguishable single photons. Photon pairs produced in bulk crystals are highly correlated in terms of transverse space and frequency. These correlations limit the indistinguishability of photons and result in inefficient photon sources.  Domain-engineered crystals with a Gaussian nonlinear response have been explored to minimize spectral correlations. Here, we study the impact of such domain engineering on spatial correlations of generated photons. We show that crystals with a Gaussian nonlinear response reduce the spatial correlations between photons. However, the Gaussian nonlinear response is not sufficient to fully eliminate the spatial correlations. Therefore, the development of a comprehensive method to minimize these correlations remains an open challenge. Our solution to this problem involves simultaneous engineering of the pump beam and crystal. We achieve purity of single-photon state up to 99 \% without any spatial filtering. Our findings provide valuable insights into the spatial waveform generated in structured SPDC crystals, with implications for applications such as boson sampling.
\end{abstract}

\pacs{Valid PACS appear here}
\maketitle

\section{Introduction}
The last years have seen an increased interest in the potential applications of quantum-based technologies. Particularly noteworthy examples are measurement-based quantum computing \cite{PhysRevA.68.022312,Walther2005}, photonic boson sampling \cite{PhysRevA.101.063821,boson}
and photonic quantum repeaters \cite{Azuma2015}. All these protocols require a sufficient amount of perfectly indistinguishable and pure single photons. Here, indistinguishability refers to the identical quantum properties of photons, while purity quantifies the degree to which a single photon is isolated from the environment. A highly pure single photon exhibits minimal correlation with its surroundings, which ensures its integrity as an independent quantum entity.

A reliable source for single photons is the nonlinear process of spontaneous parametric down-conversion (SPDC). In SPDC, a nonlinear crystal converts the high-energy photons of a laser beam into photon pairs, also known as the \textit{signal} and the \textit{idler}. Spontaneous parametric down-conversion offers two significant advantages as a single photon source. First, it is relatively simple to implement in experiments. Second, the detection of the signal photon indicates the presence of the idler photon (heralding). However, SPDC in common bulk crystals is limited by its probabilistic nature. Not all photon pairs are generated in the same state; they may be distinguishable. Moreover, the two photons are correlated with each other in terms of transverse space \cite{WALBORN201087,PhysRevA.83.033816,PhysRevA.101.043844} and frequency \cite{Steinlechner:14}. Suppose a measurement is performed only on a single photon, without consideration of the entire composite system. In that scenario, the measured photon will be in a mixed state due to the correlation with the second photon. Here, we refer to the reduced mixed state of a single photon as a single-photon state. We also distinguish between the spatial and the spectral purity of the single-photon state, which are associated with spatial and spectral correlations between signal and idler photons. The single photon is denoted as \textit{pure} if no correlation exists with its pair photon.

The most straightforward approach for elimination of the correlation between signal and idler photons is to employ spatial and spectral filtering techniques \cite{PhysRevA.101.063821}. Spatial filtering is achieved by the projection of the state into single-mode fibers (SMFs) that accept only photons in a Gaussian mode. Spectral filtering, on the other hand, involves the use of narrowband filters to select specific wavelengths for the signal and idler photons.

Although filtering can ensure high purity of the single-photon state, it is associated with high optical losses. Recent studies have explored that domain-engineered crystals can effectively minimize the spectral correlations between signal and idler photons \cite{Branczyk:11,BenDixon:13,PhysRevA.93.013801,Tambasco:16,Graffitti_2017,Graffitti:18,PhysRevApplied.12.034059,Pickston:21} without using spectral filtering. Domain-engineered crystal with Gaussian nonlinear response can deliver spectral purity of about $99$ \%. However, these studies have only considered the spectral properties of photons, where the photons have still been spatially filtered to be in a single Gaussian mode.

Until the present, no study has specifically focused on the impact of domain engineering on the spatial properties of generated photons in SPDC. We address the question if spatially pure and indistinguishable single photons can be generated from domain-engineered crystals. This can eliminate the need for spatial filtering of photons. We will follow three steps in this work, in order to answer that particular question. First, we will analyze how a Gaussian nonlinear response affects the spatial properties of photons and how it improves the spatial purity of the single-photon state. Second, we will show that another nonlinear response exists, which improves the spatial purity of the single-photon state in comparison to the Gaussian nonlinear response. In the final step, we will modify the spatial distribution of the pump beam, in order to further enhance the spatial purity. The knowledge gained in this study will take us a step closer to generating spatiospectral pure photons from SPDC without any filtering.
\section{Theory and Results}
We consider the following assumptions and approximations throughout this work: All interacting beams are assumed to be paraxial since typical
optical setups support only paraxial rays; we consider only the scalar fields instead of vector fields, since the polarization of the interacting beams is typically fixed and remains constant in the experimental realizations of SPDC; the pump beam propagates along the $z$ axis and is focused in the middle of the nonlinear crystal placed at $z=0$; signal and idler beams propagate close to the pump direction, known as the quasi-collinear regime; the narrowband approximation is applied, so only the central frequencies are generated that fulfill the energy conservation $\omega_{p}=\omega_{s}+\omega_{i}$; the transverse extension of the crystal is much larger than the pump beam waist. These assumptions and their impact on the biphoton state are discussed in more detail in our previous works \cite{PhysRevA.106.063711,sevillagutiérrez2022spectral}.

By applying these assumptions and approximations, the state of photon pairs, also known as the spatial biphoton state, obtains a simple expression \cite{PhysRevA.57.3123,PhysRevA.62.043816}
 \begin{align}\label{sinc}
    \ket{\Psi} = \iint  d\bm{q}_s \: d\bm{q}_i\:\Phi(\bm{q}_s,\bm{q}_i) \;\hat{a}^{\dagger}_s(\bm{q}_s)\:\hat{a}^{\dagger}_i(\bm{q}_i)\ket{vac}.
\end{align}
Here, $\bm{q}_{s,i}$ represent the transverse components of wave vectors of signal and idler photons, and $\hat{a}^{\dagger}_{s,i}(\bm{q}_{s,i})$ are the corresponding creation operators, respectively. The biphoton mode function function is given by \cite{WALBORN201087,Karan_2020}
\begin{equation}
    \Phi(\bm{q}_s,\bm{q}_i)= N_0\mathrm{V}_p(\bm{q}_s+\bm{q}_i)\:  \int_{-L/2}^{L/2} dz\:\chi^{(2)}(z)\,\exp{\biggr(i\Delta k_zz\biggl)}\label{modefunction},
\end{equation}
which contains the high-dimensional spatial structure of SPDC. In Eq. \eqref{modefunction}, $N_0$ is the normalization factor, $L$ is the crystal length, $\mathrm{V}_p$ is the spatial distribution of the pump beam, $\chi^{(2)}$ is the effective second-order susceptibility of the crystal, and $\Delta k_z$ is the phase mismatch in the $z$ direction. The $\Delta k_z$ reads, in the cylindrical coordinates $\bm{q}=(\rho,\varphi)$ \cite{PhysRevA.106.063711}
\begin{equation}\label{phaseMatching}
\Delta k_z=\rho_{s}^2\frac{k_p-k_s}{2k_pk_s}+\rho_{i}^2\frac{k_p-k_i}{2k_pk_i}
-\frac{\rho_{s}\rho_{i}}{k_p}\cos{(\varphi_i-\varphi_s),}
\end{equation}
where $k$ is the momentum vector $k_j=n_j\,\omega_j/c$, $n$ is the refractive index, and $c$ is the speed of light in vacuum. The pump is modeled by a common Gaussian beam profile
\begin{eqnarray*}
\mathrm{V}_p(\bm{q}_s+\bm{q}_i) \: = \: \frac{w_p}{\sqrt{2 \pi}}\:\exp{\biggl(-\frac{w_p^2}{4}|\bm{q}_s+\bm{q}_i|^2\biggr),}
  \end{eqnarray*}
with a beam waist $w_p$, unless stated otherwise.
 
Let us now consider the mathematical framework for calculation of the spatial purity. Assume, we are interested only in the idler photon and the signal photon is lost. Consequently, we need to calculate the \textit{average} outcome for each measurement on the idler photon by summing over all possible states of the signal photon. Mathematically, the averaging over the signal photon is obtained by tracing out the signal photon state from the joint state
\begin{eqnarray}
    \rho_{\mathrm{idler}}=\mathrm{Tr}_{\mathrm{signal}}(\rho),\label{trace}
\end{eqnarray}
where the density operator of the biphoton state is $\rho=\ket{\Psi}\bra{\Psi}$. The corresponding purity $P=\mathrm{Tr}(\rho_{\mathrm{idler}}^2)$ of the reduced idler state is calculated with the expression \cite{Osorio_2008} 
\begin{align}
    P &= \int d\bm{q}_s \: d\bm{q}_i \: d\bm{q}_s' \: d\bm{q}_i' \Phi(\bm{q}_s,\bm{q}_i) \: \Phi^*(\bm{q}_s',\bm{q}_i) \nonumber \\
    & \times \Phi(\bm{q}_s',\bm{q}_i') \: \Phi^*(\bm{q}_s,\bm{q}_i')\label{purtiy},
\end{align}
where we utilized Eqs. \eqref{sinc} and \eqref{trace}. The expression \eqref{purtiy} remains the same, if we switch the idler and signal photons, since $\mathrm{Tr}(\rho_{\mathrm{idler}}^2)=\mathrm{Tr}(\rho_{\mathrm{signal}}^2)$. In general, the purity of the single-photon state is always equal to or less than one. The purity $P=1$ heralds that there is no entanglement between signal and idler photons, i.e., the biphoton state can be written as a product state $\ket{\Psi}=\ket{\Psi}_s\ket{\Psi}_i$. In this regard, the purity of the single-photon state is closely related to the Schmidt number $K=1/P$ \cite{computing}, which quantifies the amount of entanglement between signal and idler photons. 
\subsection{Optimal parameters for high purity}
The crystal and pump parameters should be chosen carefully, in order to achieve high purity for the single-photon state. Here, we will mainly analyze the dependence of the purity on two parameters, the crystal length $L$ and the beam waist $w_p$. We have used potassium titanyl phosphate (KTP) crystal phase matched for type-II SPDC in our calculations. The pump laser operates at wavelengths close to $\lambda_p=775$ nm. This process produces two orthogonally polarized photons with a central wavelength of $1550$ nm. Note that the results presented in this work exhibit minimal variations across all phase-matching conditions. This reason is that the dispersion relations of different phase-matching conditions have a weaker impact on the spatial degrees of freedom (DOFs) compared to their influence on the spectral DOFs.

Initially, we will examine the spatial purity of the single-photon state in a standard periodically poled KTP crystal. The presented results are also valid for bulk crystals, producing quasi-collinear SPDC. Subsequently, we assess the performance of KTP and bulk crystals relative to domain-engineered crystals with a Gaussian and also with a more general nonlinear response.

\subsubsection{Periodically poled crystal}
Bulk and periodically poled crystals possess a $sinc$-shaped phase-matching function,
\begin{equation}
    \Phi(\bm{q}_s,\bm{q}_i)= N_0\mathrm{V}_p(\bm{q}_s+\bm{q}_i)\:  L\sinc(L\Delta k_z/2).\label{sinclike}
\end{equation}
We can directly insert the expression \eqref{sinclike} into Eq. \eqref{purtiy} and calculate the purity of the single-photon state. Alternatively, we can keep our calculations more general and insert the expression \eqref{modefunction} into Eq. \eqref{purtiy} and perform the integrals for the general $\chi^{(2)}$. The eight integrals over the transverse momentum can be solved analytically, while the four integrals over $z$ need to be performed numerically. The sinclike behavior of the phase-matching is reconstructed by either inserting a  periodically changing function for $\chi^{(2)}$ into Eq. \eqref{modefunction} or by inserting $\chi^{(2)}=1$ for bulk crystals. Both lead to the same result.

\begin{figure}[t!]
\includegraphics[width=.47\textwidth]{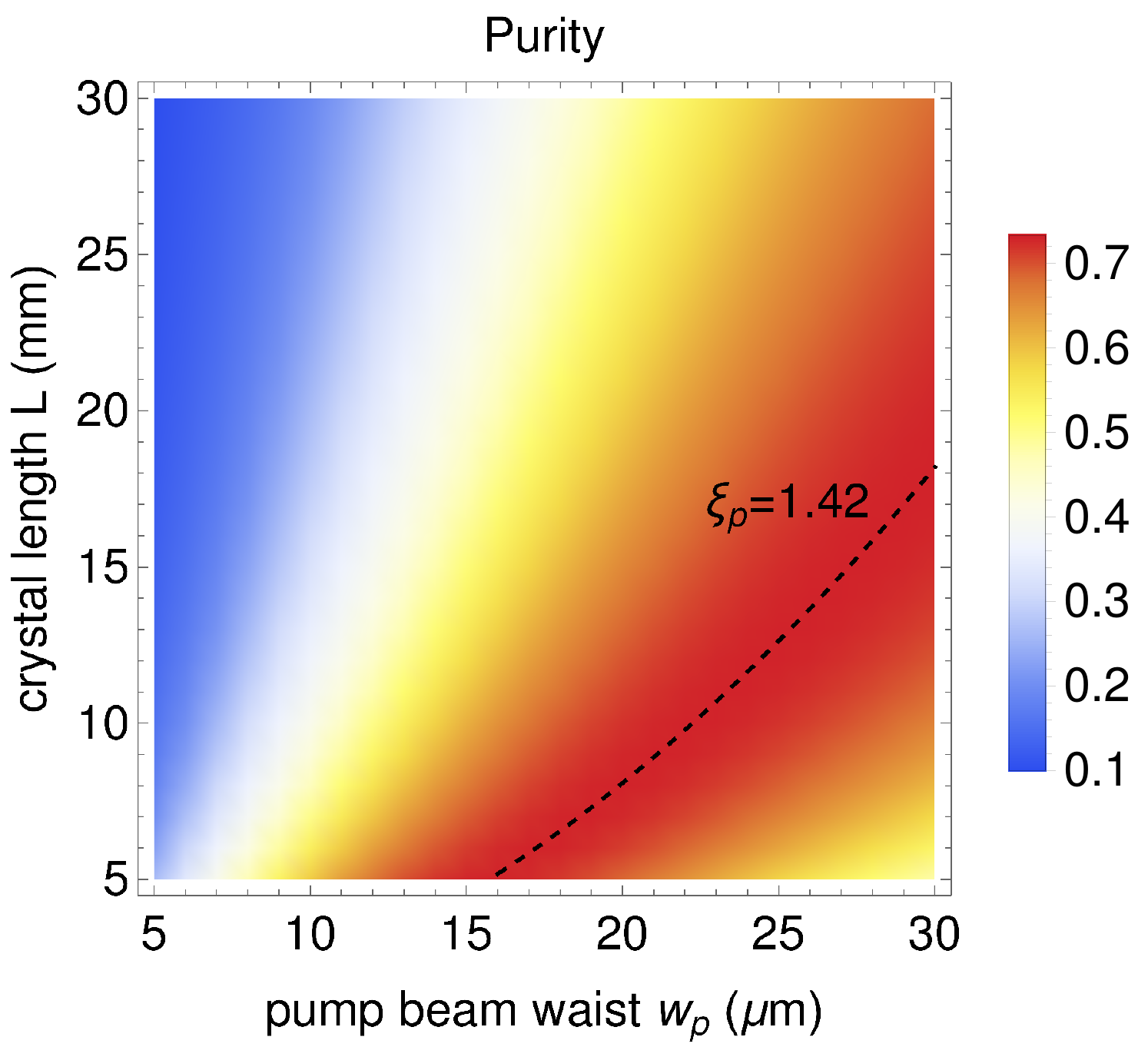}
\caption{Spatial purity as a function of the pump beam waist $w_p$ and crystal length $L$ for a standard periodically poled or bulk crystal. The purity reaches its maximum $P=0.73$ along the black dashed curve, corresponding to the beam parameter $\xi_p=1.42$.}\label{wLsinc}
\end{figure}

Figure \ref{wLsinc} shows the spatial purity as a function of the beam waist $w_p$ and the crystal length $L$. Notably, the purity remains approximately constant along the curves $L\propto w_p^2$. This is not surprising, since the spatial biphoton state should depend only on the dimensionless focusing parameter $\overline{w}_p=w_p/\sqrt{\lambda_p\,L}$ already discussed in Refs. \cite{PhysRevA.68.050301,PhysRevA.103.063508} or, equivalently, the beam parameter $\xi_p=L/(k_p\, w_p^2)$ discussed in Ref. \cite{Palacios:11} assuming the degeneracy condition $k_s\approx k_i\approx k_p/2$. The highest possible purity equals $0.73$, which remains constant on the black dashed curve corresponding to the beam parameter $\xi_p\approx 1.42$. This implies that we have the freedom to choose any pair $(w_p,L)$ on this curve.

Most SPDC experiments include SMFs, where the higher the pair collection probability, the more efficient the setup. Therefore, the choice of the proper pair $(w_p,L)$ along the curve can be based on the optimization of the pair collection probability into the SMF. The projection modes of the SMF are known to be approximately Gaussian \cite{PhysRevA.72.062301}
\begin{eqnarray*}
    U(\bm{q},w)= \: \frac{w}{\sqrt{2 \pi}}\:\exp{\biggl(-\frac{w^2}{4}|\bm{q}|^2\biggr)}.
  \end{eqnarray*}
Therefore, the pair collection probability into the SMF is given by the expression
\begin{equation}
  R^{(2)} = \abs{\iint d\bm{q}_s \: d\bm{q}_i \Phi(\bm{q}_s,\bm{q}_i) [U(\bm{q}_s,w_s)]^* [U(\bm{q}_i,w_i) ]^*}^2,\label{PCP}
\end{equation}
where $w_{s}$ and $w_{i}$ are the collection waists of the signal and idler beams, respectively.
\begin{figure}[t!]
\includegraphics[width=.47\textwidth]{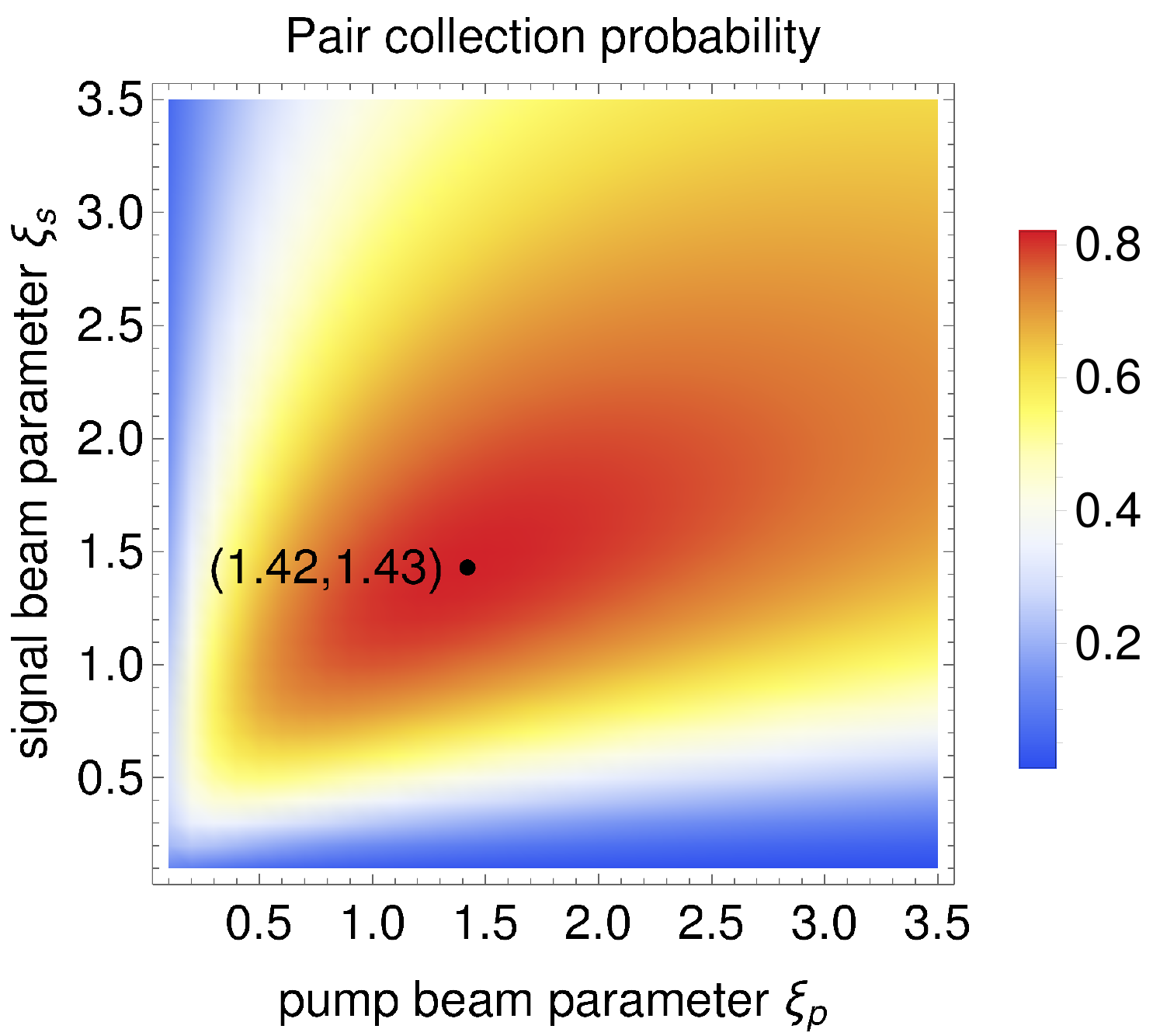}
\caption{Pair collection probability into the SMF as a function of the pump $\xi_p$ and signal $\xi_s$ beam parameters, where the condition $\xi_s=\xi_i$ is assumed. The highest pair collection probability is achieved at the same beam parameter $\xi_p=1.42$, which simultaneously maximizes the purity.}\label{pairCollection}
\end{figure}
The integrals in Eq. \eqref{PCP} can be solved analytically for the degenerate scenario $k_s=k_i=k_p/2$ \cite{Palacios:11}. The authors of Ref. \cite{Palacios:11} have shown that the highest coupling efficiency is achieved for the beam parameter $\xi_s=\xi_i=\xi_p=1.39$ with a pair collection probability into the SMF of $R^{(2)}\approx 82$\%. Figure \ref{pairCollection} shows the pair collection probability as a function of pump and signal (idler) beam parameters.  We calculated numerically similar values of $\xi_p\approx 1.42$ and $\xi_s=\xi_i\approx 1.43$, which lead to $ R^{(2)}\approx 82.2$ \%. Due to birefringence in the type-II quasi-phase-matching configuration considered in our work, these parameters deviate slightly from the results of Ref. \cite{Palacios:11}, which were obtained under the assumption of degeneracy, $k_s=k_i=k_p/2$.

Remarkably, the optimal values of $\xi_p$ required to achieve maximum purity and pair collection probability into the SMF are identical: The optimization of the purity is equivalent to the optimization of the pair collection probability into the SMF. Here is an explanation for this observation: By enhancing the purity, we force the signal and idler photons to be more $ concentrated $ in a specific mode. Consequently, this reduces the spread of the state across different modes. In an ideal scenario, where $P=1$, each photon occupies a single-mode state, denoted as $\ket{\Psi}=\ket{\Psi}_s\ket{\Psi}_i$ (a product state). It turns out that this single-mode state $\ket{\Psi}_{s,i}$ corresponds to a Gaussian state, when we try to enhance the spatial purity. Therefore, the pair collection probability into the SMF, which accepts only photons in a Gaussian mode, increases. However, we will see in Sec. \ref{pumpEngineering} that the single mode, where the state is mostly concentrated, is the Gaussian state, only if the pump is also described by a Gaussian.

\subsubsection{Gaussian nonlinear response}
We demonstrate in this section that the spatial purity of the single-photon state and pair collection probability into the SMF can be significantly enhanced by using domain-engineered crystals. We first employ a Gaussian nonlinear response and investigate its impact on the purity of the single-photon state. In the next section, we will employ an optimized nonlinear response and compare its performance with the Gaussian nonlinear response.

A domain-engineered crystal with a Gaussian nonlinear response can be described by $\chi^{(2)}\propto \exp{[-z^2/(\sigma)^2]}$ (see Fig. \ref{fig3}). The behavior of the spatial purity for this state as a function of $w_p$ and $L$ is similar to Fig. \ref{wLsinc}. The purity of the single-photon state remains approximately constant on curves corresponding to $L\propto w_p^2$ or equivalently to $\xi_p=const.$. 

Here, we vary two parameters, namely, the width $\sigma$ of the Gaussian and the beam parameter $\xi_p$, in order to achieve high purity. The maximum possible value of the spatial purity can be increased up to $P=0.95$ for the parameters $\xi_p\approx 3$ and $\sigma =L/4$. The pair collection probability into the SMF is enhanced up to $R^{(2)}\approx97$ \%. As we see, domain engineering has the same impact on the spatial properties of photons as it has on the spectral properties: Domain-engineered crystals with a Gaussian nonlinear response can reduce the spatial correlations enormously (high spatial purity). However, the spectral purity $P=0.99$ achieved in the previous works is still not accessible for the spatial DOFs of photons.

We might expect to achieve higher purity if a narrower Gaussian function for the susceptibility ($\sigma < L/4$) is used. This is because a narrower Gaussian function generates fewer modes, which might result in a higher spatial purity of the single-photon state. Surprisingly, the purity can not be improved by making $\sigma$ smaller. While a Gaussian function with a narrower width works well for the generation of spectrally pure single-photon states, a more general function is needed to improve the spatial purity of the single-photon state. The reason is that the phase mismatch \eqref{phaseMatching} in the $z$ direction of the spatial biphoton state is more complicated than the phase mismatch of spectral biphoton state \cite{PhysRevA.106.063714}.

\subsection{Crystal engineering}\label{crystal}
In this section, we try to find a more general nonlinear response that delivers higher purity than the Gaussian nonlinear response.  We expect the optimal susceptibility function $\chi^{(2)}(z)$ to be symmetric to the axis $z=0$, since the well-working Gaussian function is symmetric to that axis too. Therefore, we expand $\chi^{(2)}(z)$ into cosine series, similar to Fourier series for even functions: 
\begin{equation}
    \chi^{(2)}(z)=\sum_{n=0}^N c_n \cos{(n z/\sigma)},\label{cosine}
\end{equation}
where the expansion coefficients $c_n$ are initially unknown and $\sigma= L/4$. The purity as a function of $c_n$, $ P=P(c_0,c_1,...,c_N)$, can be constructed by inserting Eq. \eqref{cosine} into Eq. \eqref{purtiy}. We truncate the sum at $N=7$ in our analysis. The following steps are performed, in order to find the optimal expansion coefficients $c_n$. (i) We construct the function $P=P(c_0,c_1,...,c_7)$ and calculate its local maximum for the fixed beam parameter $\xi_p=1.42$ from the preceding section. The function \textit{FindMaximum} from Wolfram Mathematica v. 12.3 \citep{Mathematica} is used to calculate the local maximum of the function $ P=P(c_0,c_1,...,c_7)$. (ii) We construct the susceptibility function $\chi^{(2)}(z)$ based on the calculated coefficients $c_0,c_1,...,c_7$. (iii) We optimize the purity as a function of the beam parameter, where the reconstructed susceptibility function $\chi^{(2)}(z)$ is used. (iv) We repeat the first three steps for the replaced optimal beam parameter untill the purity converges. The obtained results are independent of $\sigma$, which just determines how fast the cosine series converge.

\begin{figure}[t!]
\includegraphics[width=.47\textwidth]{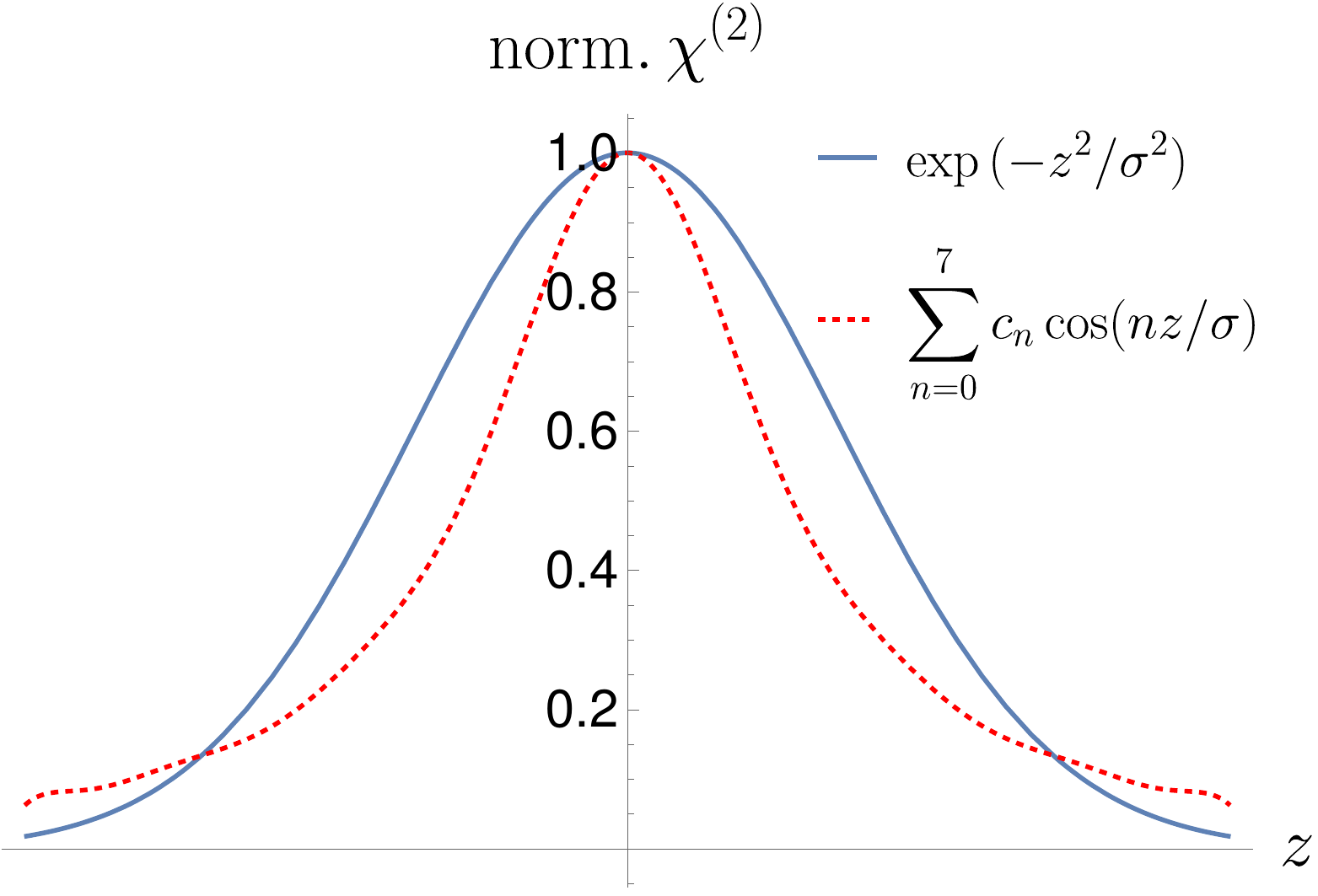}
\caption{Comparison of normalized Gaussian nonlinear response (blue solid line) with cosine series (red dashed line) susceptibility function $\chi^{(2)}$. The latter ensures higher purity for the single-photon state compared to the Gaussian response.}
\label{fig3}
\end{figure}

\begin{figure}[t!]
\includegraphics[width=.47\textwidth]{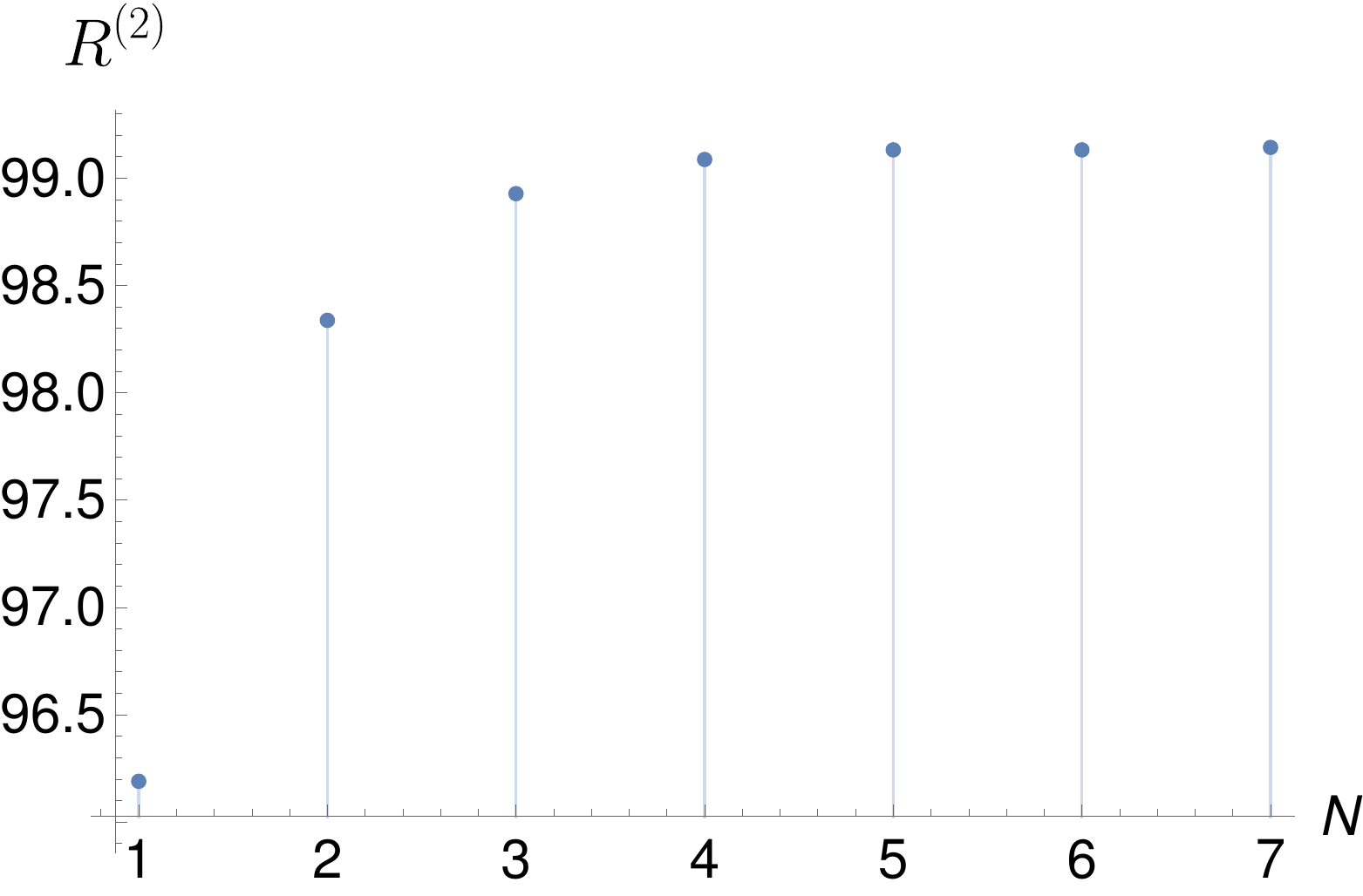}
\caption{Pair collection probability into the SMF, which depends on the number of terms in the cosine series for $\chi^{(2)}(z)$. We see that even with a relatively small number of terms, such as $N=5$, we can achieve a high collection probability of $99\%$. }
\label{cosineFig}
\end{figure}

The final optimal value for the beam parameter is equal to $\xi_p=3.67$, and the corresponding $c_n$ coefficients are $-0.2904$, $0.6799$, $-0.4851$, $0.3903$, $-0.2195$, $ 0.1242$, $-0.0440$, $0.01487$ for $n=0,1,2,3,4,5,6$ and $7$, respectively. The calculated \textit{cosine}-series and the Gaussian susceptibility functions $\chi^{(2)}$ are displayed in Fig. \ref{fig3} for comparison. The optimized susceptibility function $\chi^{(2)}(z)$ indeed enhances the purity up to $0.98$ and the pair collection probability into the SMF up to $R^{(2)}\approx99$ \%. This shows that the Gaussian nonlinear response is not the best choice, to achieve high spatial purity for the single-photon state. In order also to show how fast this cosine series converges, we calculate the pair collection probability as a function of the number of terms $N$ in the cosine series displayed in Fig. \ref{cosineFig}.

In Table \ref{tabele}, we summarize the performances of crystals with different nonlinear responses in regard to spatial purity, pair collection probability into the SMF, and heralding efficiency. The heralding efficiency of the single photon is obtained by tracing out the possible wave vectors of the partner photon (here idler)
\begin{equation*}
  \eta = \int d\bm{q}_i\abs{\int d\bm{q}_s \: \Phi(\bm{q}_s,\bm{q}_i) [U(\bm{q}_s,w_s) ]^*}^2.
\end{equation*}
It quantifies the probability that the detection of a signal photon in a Gaussian mode will lead to the successful detection of its entangled partner.
\begin{table}
\begin{ruledtabular}
\begin{tabular}{ccccc}
  Parameter& Sinc & Gaussian & Cosine\\ \hline
 purity of the reduced state &$0.73$&$0.95$ &$0.98$ \\
 coupling into the SMF &$82\%$
 &$97\%$&$99\%$\\
Heralding efficiency &$99.4\%$
 &$99.86\%$&$99.98\%$\\
\end{tabular}
\end{ruledtabular}\caption{Comparison of performances of different nonlinear responses in terms of purity, pair collection probability into a SMF, and heralding efficiency. The considered nonlinear responses are the common sinclike behavior from periodically poled or bulk crystals, and the Gaussian and cosine series from domain-engineered crystals.}\label{tabele}
\end{table}

\subsection{Custom-poling}
The effective nonlinearity of the crystal accessible in the experiment is encoded in the phase-matching function. The phase-matching function of a crystal with susceptibility given by Eq. \eqref{cosine} reads
\begin{align}
      \label{phaseMatching3}
    \phi(\Delta k_z)&= \int_{-L/2}^{L/2} dz\:\sum_{n=0}^7 c_n \cos{(n z/\sigma)}\,\exp{\biggr[i\Delta k_zz\biggl]}\nonumber\\&
    =\frac{L}{2}\sum_{n=0}^7 c_n\,[\sinc(2n-\Delta k_zL/2)\nonumber\\&+\sinc(2n+\Delta k_zL/2)].
\end{align}

In this theoretical treatment, the susceptibility $\chi^{(2)}$ is modeled as a continuous function of $z$. However, in practice, the susceptibility in a crystal typically modulates in a discontinuous manner, via discrete domain inversion of the nonlinear susceptibility. Specifically, this results in a poling function that only takes on values of $\pm \chi^{(2)}_0$, where $\chi^{(2)}_0$ is the modulus of the susceptibility $\chi^{(2)}$. To account for this, we employ a domain-engineering process, wherein the nonlinear medium consists of $M$ domains of birefringent material, each with a length of $l_c$. These domains can be oriented in either the \textit{up} or the \textit{down} direction. Consequently, the phase-matching function for the entire crystal becomes a linear superposition of individual phase-matching functions for each crystal domain \cite{PhysRevA.93.013801}
\begin{equation*}
    \chi^{(2)}(z)=\chi^{(2)}_0\sum_{m=1}^M s_m\, \mathrm{rect}\biggl(\frac{z-z_m}{l_c}\biggl),
\end{equation*}
where $s_m$ takes on a value of either $1$ or $-1$ depending on the orientation of the domain, and $z_m=(m - 1/2)l_c-L/2$ specifies the positions of the $m$th domain. 

\begin{figure}[t!]
\includegraphics[width=.47\textwidth]{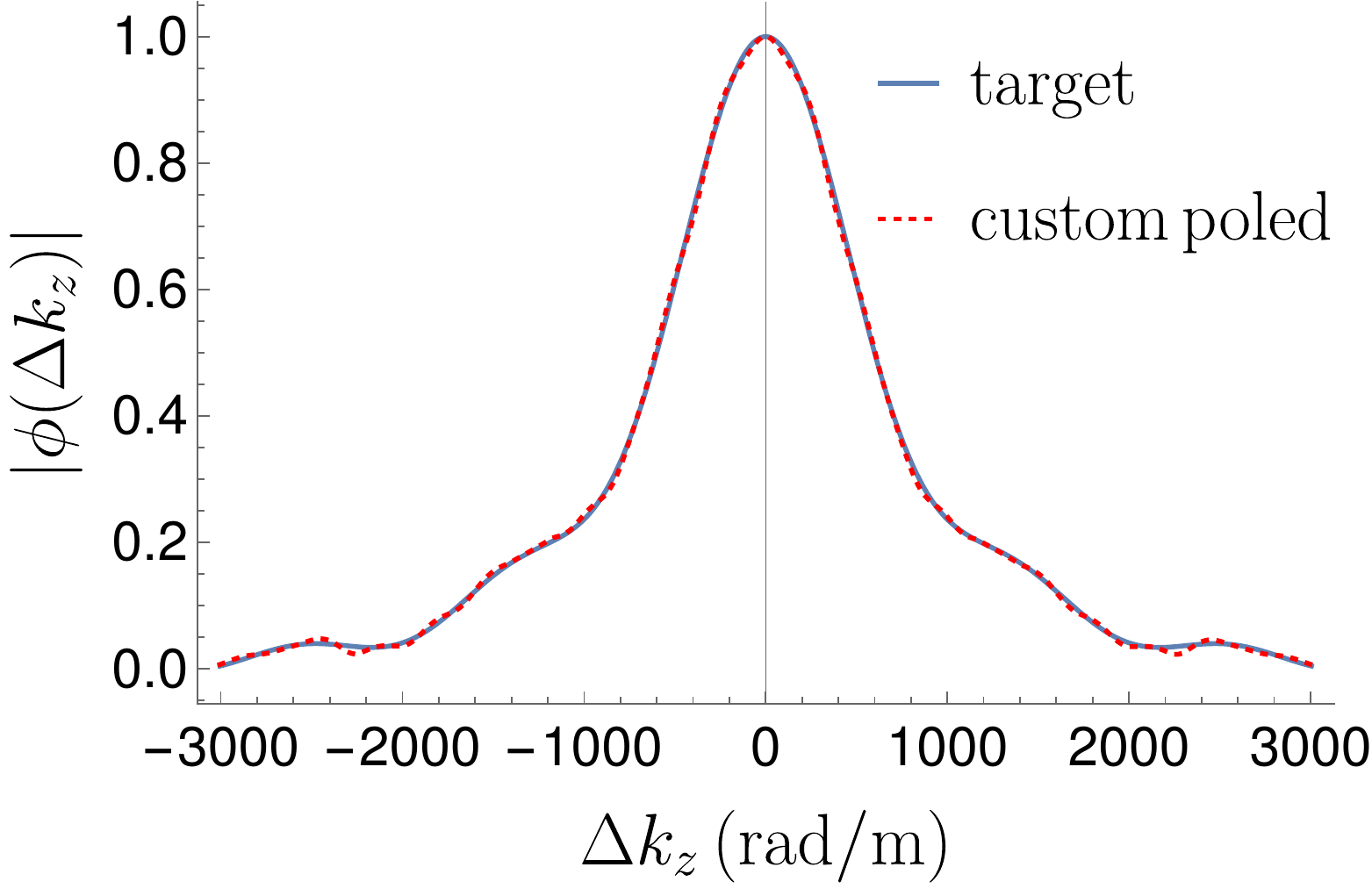}
\caption{Comparison of the normalized target phase-matching function calculated by Eq. \eqref{phaseMatching3} (blue solid line) with the phase-matching function of a custom poled crystal (red dashed line). Indeed, the nontrivial target function can be very well designed by using custom-poled crystals.}
\label{fig5}
\end{figure}

In order to engineer the susceptibility functions $\chi^{(2)}$ shown in Fig. \ref{fig3}, the correct sequence $s_m$ of the relative orientation of domains needs to be determined. Here, we use the code provided by Dosseva \textit{et} \textit{al}. \cite{PhysRevA.93.013801}, to find the customized poling of the crystal for the cosine series $\chi^{(2)}$. The blue solid line in Fig. \ref{fig5} shows the target phase-matching function given by Eq. \eqref{phaseMatching3} and the red dashed line represents the phase-matching function of the custom poled crystal obtained through the code from Ref. \cite{PhysRevA.93.013801}. Our calculations employ $1300$ domains, each with a length of $23$ $\mu$m, resulting in a total crystal length of $L=29.9$ mm. As demonstrated in Fig. \ref{fig5}, the complex phase-matching function described by Eq. \eqref{phaseMatching3} can be excellently reproduced, confirming its applicability in real experimental setups.

Except for the code provided in Ref. \cite{PhysRevA.93.013801}, many other groups further developed algorithms to calculate the optimal domain configuration for desired nonlinearity. For example, more recent algorithms that can enhance the accuracy of domain engineering can be found in Refs. \cite{Graffitti_2017,PhysRevApplied.12.034059}.

Overall, the effective nonlinearity of an engineered crystal is lower than that of a periodically poled crystal of the same length. However, in many experimental settings, the enhancement in purity offsets the cost of reduced efficiency. Moreover, it is usually possible to increase the power of the laser or use a longer crystal, in order to maintain the pair generation rate.

A natural next step for further investigation would be to include both spectral and spatial DOFs of photons in the optimization scheme of the nonlinear response, in other words, to design a source of pure spatiospectral photon pairs from SPDC that requires no spectral or spatial filtering. The best purity we achieve here using optimized nonlinear response, $P=0.98$, is still quite far from unity and the consideration of the spectral DOF would further decrease this value. Therefore, achieving spatiospectral purity close to unity with only domain-engineered crystals seems to be a very challenging task. 
\subsection{Pump engineering}\label{pumpEngineering}
In this section, we address the question if the engineering of the pump can enhance the spatial purity of the single-photon state. The engineering of the pump has been already used to control the spatial correlation in SPDC \cite{PhysRevA.98.060301,PhysRevA.102.052412,PhysRevA.106.063711,Boucher:21,Rozenberg:22}.
Here, pump engineering is similar to the crystal engineering method. We make an ansatz of the superposition of Laguerre-Gaussian (LG) beams for the pump
\begin{eqnarray}
   \mathrm{V}_p & = & \sum_{p,\ell}a_{p,\ell}\;\mathrm{LG}_{p}^{\ell}.\label{pump}
\end{eqnarray}
The mode numbers of the LG beam $p$ and $\ell$ are associated with the radial momentum and projection of the orbital angular momentum of photons, respectively. The method is the same: Find the best set of coefficients $a_{p,\ell}$ which maximize the spatial purity. This is a more difficult task than crystal engineering, since the eight-dimensional integral in Eq. \eqref{purtiy} should be solved fully numerically if the pump is given by Eq. \eqref{pump}. This calculation requires high numerical accuracy, which makes it computationally expensive. Therefore, we need to simplify the task to some extent.

We derived a general expression for the spatiospectral biphoton state pumped by a LG beam in our previous work \cite{PhysRevA.106.063711}. In addition, the state was decomposed into LG modes as 
\begin{align}\label{decomposition}
    \ket{\Psi}= \sum_{p_s,p_i=0}^{\infty}\: \sum^{\infty}_{\ell_s,\ell_i=-\infty}\;C_{p,p_s,p_i}^{\ell,\ell_s,\ell_i} \ket{p_s,\ell_s}\ket{p_i,\ell_i},
\end{align}
where $\ell_{s,i}$ and $p_{s,i}$ are the mode numbers of the signal and idler photons, respectively. We omitted here the frequency dependence of the coincidence amplitudes compared to Ref. \cite{PhysRevA.106.063711} since we only consider the narrowband regime. The full expression of the coincidence amplitudes $C_{p,p_s,p_i}^{\ell,\ell_s,\ell_i}$ can be found in Ref. \cite{PhysRevA.106.063711}. If the pump beam is given by Eq. \eqref{pump}, the coincidence amplitudes are updated to a similar linear superposition, $\sum_{p,\ell}a_{p,\ell}\;C_{p,p_s,p_i}^{\ell,\ell_s,\ell_i}$. We can truncate the infinite summations in Eq. \eqref{decomposition} at reasonable mode number values and consider the subspace state instead of the full state \eqref{sinc}. This will enormously simplify our calculation. The subspace state reads then
\begin{align}
    \ket{\Psi_s}= \sum_{p_s,p_i=0}^{H}\: \sum^{U}_{\ell_s,\ell_i=-U}\;\biggl(\sum_{p,\ell}a_{p,\ell}\;C_{p,p_s,p_i}^{\ell,\ell_s,\ell_i} \biggr)\ket{p_s,\ell_s}\ket{p_i,\ell_i}.
\end{align}
where $H$ and $U$ are the boundaries of the subspace.

We use the phase matching from Sec. \ref{crystal} and construct the purity function depending on $a_{p,\ell}$ coefficients, $P=P(a_{p_1,\ell_1},a_{p_2,\ell_2},...,a_{p_n,\ell_n})$. The summation of pump beam modes is carried out over the range of $p=0,1,2$ and $\ell=-3,-2,...,3$. The highest possible purity in this range equals $P\approx0.99$. This is an improvement to the purity $P=0.98$ from the preceding section. However, the pair collection probability drops off to $0.64$ and the heralding efficiency to $99.3\%$, since the generated single mode is not the fundamental Gaussian mode anymore. 
\begin{figure}[t!]
\includegraphics[width=.47\textwidth]{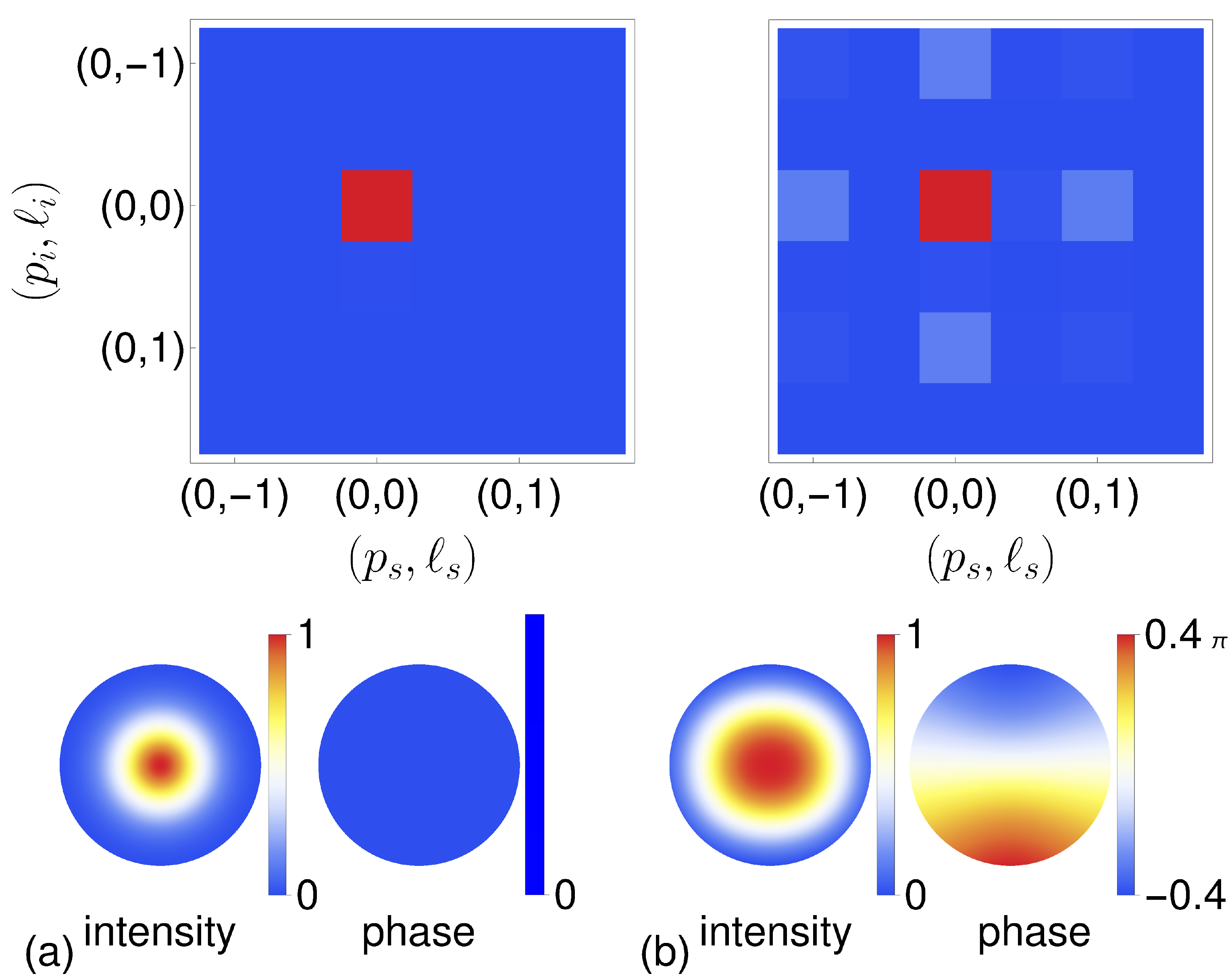}
\caption{Mode distribution of the biphoton state in the LG basis for (a) a Gaussian and (b) for an engineered pump beam. The corresponding beam profiles and the phase distributions are presented at the bottom. The calculation has been performed for the phase-matching function from Sec. \ref{crystal}.}
\label{fig6}
\end{figure}

Figure \ref{fig6} shows the decomposition of signal and idler photons in the LG basis (correlation matrix) for the Gaussian and the engineered pump. The engineered pump symmetrizes the mode distributions but at the cost of the Gaussian mode. While the generated single mode in the case of a Gaussian pump is approximately the Gaussian mode, the single mode generated by the engineered pump is now a superposition of LG modes. This is the reason, why the pair collection probability into the SMF is decreased in the case of the engineered pump.

This kind of source might be interesting for experiments which are sensitive not to the pair collection probability into the SMF but rather to the purity of photons in some interference experiments.

\subsection{Efficiency of pump engineering}
We showed that a shaped pump beam can enhance the spatial purity. On the other hand, it is natural to ask if a shaped pump beam decreases the pair creation probability compared to Gaussian beam. In general, the efficiency of pair generation will not suffer from the change of the pump from Gaussian to the engineered pump too much if the intensity of the laser remains the same. This statement can be proved by calculating the total pair collection probability
 \begin{equation}
   R_{total}= \iint  d\bm{q}_s \: d\bm{q}_i\:\abs{\Phi(\bm{q}_s,\bm{q}_i)}^2.\label{total}
\end{equation}
These integrals are very easy to solve in the degenerate scenario $k_s=k_i=k_p/2$. If we use the notation $\bm{q}_{-}=\bm{q}_s-\bm{q}_i$ and $\bm{q}_{+}=\bm{q}_s+\bm{q}_i$, the biphoton mode function in the degenerate scenario becomes
\begin{eqnarray*}
        \Phi(\bm{q}_{+},\bm{q}_{-})= N_0\mathrm{V}_p(\bm{q}_{+})\: \phi(\bm{q}_{-}),
\end{eqnarray*}
where the phase-matching function $\phi(\bm{q}_{-})$ reads
\begin{equation*}
 \phi(\bm{q}_{-})=   \int_{-L/2}^{L/2} dz\:\chi^{(2)}(z)\,\exp{\biggr[i\frac{\abs{\bm{q}_{-}}^2}{2k_p}z\biggl]}.
\end{equation*}
The integrals in Eq. \eqref{total} can then be simplified according to Ref. \cite{PhysRevA.106.063714},
 \begin{equation}
   R_{total}= \frac{1}{4}N_0^2\:\int d\bm{q}_{-}\, \abs{\phi(\bm{q}_{-})}^2,\label{total2}
\end{equation}
where it is assumed that the pump profile is normalized $\int d\bm{q}_{+} \abs{\mathrm{V}(\bm{q}_{+})}^2=1$. It follows from Eq. \eqref{total2} that the total rate is independent of the pump beam profile as long as the intensity of the laser remains constant. The total rate becomes dependent on the pump beam profile only if the crystal produces nondegenerate SPDC. Nevertheless, based on our calculations, this dependence is relatively weak.

Note that in this analysis, we have not considered the possible losses arising from the experimental realization of transforming the Gaussian to an engineered pump beam profile. Recent beam shaping approaches, such as multiplane light conversion techniques using custom computer-generated holograms \cite{Fontaine2019,Hiekkamaki:19}, can transform spatial modes with, in principle, unit efficiency.

\section{Conclusion}
In this work, we studied the impact of domain engineering on the spatial properties of SDPC. We answered the question of how the spatial purity of the single-photon state can be improved by using domain-engineered nonlinear crystals. We found that a more general nonlinear response rather than the Gaussian one exists that improves the spatial purity of the single-photon state, pair collection probability into the SMF, and heralding efficiency. We also showed that the spatial purity can be enhanced further by engineering the pump beam, but at the cost of the pair collection probability into the SMF.

Overall, it is not obvious, that the generation of spatially pure single photons from domain-engineered crystals is more efficient than the spatially filtered photons produced in bulk crystals. However, when it comes to generating spectrally pure photons, the use of engineered crystals becomes the preferred choice. This preference is driven by the significant challenges associated with achieving the desired spectral state through filtering. In general, the spectral filtering is more complex to implement than spatial filtering. In this context, when considering both spatial and spectral DOFs simultaneously, the use of filters will not be desired due to the spatiospectral coupling in SPDC. Hence, the engineered crystals will offer a more practical solution. The ultimate goal will be then the generation of spatiospectral pure photons without any spatial and spectral filtering.

Summarizing, our findings have important implications for the use of quantum-based technologies, as they eliminate the need for spatial filtering of photons and enable the generation of high-purity photons without incurring significant optical losses. Moreover, the insights acquired through this study will move us one step closer to achieving the generation of spatiospectral pure photons from SPDC without the need for any filtering.

\section*{Acknowledgement}
The authors thank Francesco Graffitti and Carlos Sevilla-Gutiérrez for helpful discussions and René Sondenheimer for his useful suggestions on how to improve the paper.

\bibliographystyle{apsrev4-1}
\bibliography{bibliography.bib}      
\end{document}